Two-dimensional gel electrophoresis in proteomics: a tutorial

Thierry Rabilloud [1,2] , Cécile Lelong [3]

1 : CNRS UMR5249, Chemistry and Biology of Metals, CEA Grenoble, iRTSV/BSBBSI, 17 rue des martyrs, F-38054 GRENOBLE CEDEX 9

2 : CEA-DSV/iRTSV/LCBM, Chemistry and Biology of Metals, CEA-Grenoble, 17 rue des martyrs, F-38054 GRENOBLE CEDEX 9, France

3 : Université Joseph Fourier, UMR CNRS-CEA-UJF 5092, CEA-Grenoble, 17 rue des martyrs, F-38054 GRENOBLE CEDEX 9

Correspondence to

Thierry Rabilloud, iRTSV/LCBM

CEA-Grenoble, 17 rue des martyrs,

F-38054 GRENOBLE CEDEX 9

Tel (33)-4-38-78-32-12

Fax (33)-4-38-78-44-99

e-mail: Thierry.Rabilloud@ cea.fr

Abstract

Two-dimensional electrophoresis of proteins has preceded, and accompanied, the birth of proteomics. Although it is no longer the only experimental scheme used in modern proteomics, it still has distinct features and advantages. The purpose of this tutorial paper is to guide the reader through the history of the field, then through the main steps of the process, from sample preparation to in-gel detection of proteins, commenting the constraints and caveats of the technique. Then the limitations and positive features of two-dimensional electrophoresis are discussed (e.g. its unique ability to separate complete proteins and its easy interfacing with immunoblotting techniques), so that the optimal type of applications of this technique in current and future proteomics can be perceived. This is illustrated by a detailed example taken from the literature and commented in detail.

## 1. Historical Background

In separation sciences, there is always a trend toward high-resolution separations, in order to be able to analyze complex samples. This is of course very strong in biology, where the samples are most of the times very complex. Electrophoretic separations of proteins are no exception to this rule, with the additional difficulty that proteins are very complex analytes that have a strong tendency to precipitate. At the very beginning of the 70's, two high-performance electrophoretic separations of proteins were available: i) zone electrophoresis of proteins in the presence of SDS, as described in its almost final form by Laemmli [1], a technique that instantly became very popular, and still is, and ii) denaturing isoelecric focusing, as described for example by Gronow and Griffith [2]. As these two techniques used completely independent separation parameters (molecular mass and isoelectric point, respectively) it is not surprising that it was soon tried to couple them. The first successful report, in 1974 [3], got almost unnoticed, because the difficult method of sample inclusion in the IEF gel was used, and because the gels were stained by the rather insensitive Coomassie blue staining, thereby showing only a few spots on the gel, thus something rather unimpressive. However, in 1975, the next report [4] completely changed the situation. It was a very detailed report, using a cathodic sample application (thus adaptable to series of samples) and showing hundreds of distinct spots through the used of autoradiography of 35S-labelled proteins. The resolution was, as expected, much greater than with other two-dimensional techniques, e.g. those using native isoelectric focusing [5] or those using another type of zone electrophoresis as the first dimension [6]. In fact, the results of the O'Farrell technique were so impressive that many protein biochemists took on that technique, and the first premise of what is now known as the Human Proteome Project traces back to these early years [7]. However, two core features of these early days had very strong consequences. The first one is the rather poor reproducibility of isoelectric focusing with carrier ampholytes, which is prone to several problems such as cathodic drift [8]. This meant that it was very difficult to achieve good run-to-run reproducibility. Consequently, methods allowing to run several gels in parallel were developed to increase sample to sample reproducibility [9, 10], and this parallelicity is still very widely used today.

The second one was the absence of techniques enabling to identify a specific protein spot on a 2D gel. For example, it took several years of hard work to demonstrate that PCNA, early identified as a protein spot modulated by proliferation [11], was in fact a DNA polymerase subunit [12]. Thus, the best way to utilize the wealth of data present in a series of 2D gels was to perform large-scale data analysis. This started very early after the introduction of 2D gels, [13, 14], and was soon refined to perform multivariate analyses [15-19] that are in fact rediscovered today.

These two major pitfalls were the subject of intense research that succeeded at the end of the 80's, on the one hand with the introduction of immobilized pH gradients [20-22] and on the other hand with protein identification with Edman sequencing [23-25]. With immobilized pH gradients the reproducibility of the separations made a quantum leap, allowing even interlaboratory comparison [26], and the scope of separations was also dramatically improved [27]. With protein sequencing it became possible to identify spots on 2D gels almost at will (e.g. in [28]), although it was much more labor-intensive and less sensitive with Edman sequencing than it is now with mass spectrometry, which also achieved a quantum leap in speed, sensitivity and depth of protein characterization after 2D electrophoresis.

## 2. Basic concepts

The basic concept of 2D electrophoresis is schematized in Figure 1. Although the name of 2D electrophoresis suggests that it is a two-step process, it is indeed a five-step process starting from sample preparation prior to the first separation (step 1), then first separation (step 2), then interfacing with the second separation (step 3) then second separation (step 4) then finally protein detection  (step 5). Generally, the order of separation is isoelectric focusing first and SDS electrophoresis second. However, this is not a mandatory figure, and the reverse order has also been described [29, 30]. Nevertheless, the classical order is almost exclusively used, both for economical reasons (the large, second separation is better to be the cheapest one, i.e. SDS electrophoresis) and for technical reasons, as SDS electrophoresis gels are much easier to stain than isoelectric ones, and also much easier to interface with downstream protein analysis techniques, whether this consists in protein blotting or in mass spectrometry. Both are very sensitive to the presence of ampholytes and of nonionic detergents, two types of chemicals that are almost indispensable in denaturing isoelectric focusing, but are limited to the buffer front when the classical order of separations is used.

It is quite beyond the scope of such a tutorial paper to describe detailed protocols on how to prepare samples for 2D electrophoresis, run them on 2D gels and finally detect the proteins. There is a wealth of books that provide such detailed and commented protocols. Some of them are now only of historical interest [31], but some others, cited here, are still of practical interest [32-34] or allow to go much deeper in the understanding of some aspects of 2D gels [35] . Some review and/or practical papers are also of interest, either because they describe important caveats [36], or because they go deeper into more specialized topics, such as protein detection [37], non classical (i.e. without IEF) 2D gels [38] or variations to the basic electrophoretic protocols [39].
However, within the scope of this tutorial, we would like to show "the back side of the cards" in a stepwise fashion, and help the reader to understand the key parameters at play in 2D electrophoresis and the rationale underlying the protocols.

### 2.1. Sample preparation

In an ideal world, sample preparation would solubilize quantitatively all proteins, without any modification added during the whole process, in a way fully compatible with the first separation, and would at the same eliminate all other biological compounds that might interfere with this first separation. As the first separation is always almost isoelectric focusing, this means that sample separation will be constrained by isoelectric focusing. Indeed, isoelectric focusing brings two types of constraints.

First of all, the charge of the proteins must not be altered in any way. This means in turn that the best trick that we have at hands for solubilizing proteins, i.e. the use of charged detergents such as SDS, cannot be used. Thus, we must rely on uncharged chemicals for extracting the cellular proteins, denaturing them and keeping them in solution. To achieve this purpose, a combination of chaotropes and detergents is almost always used. Chaotropes are excellent protein denaturing and solubilizing agents [40], and neutral chaotropes such as urea (since the very beginning of denaturing IEF [2]) and the more recently introduced thiourea [41] are commonly used. However, chaotropes alone are not sufficient, as they do not dissolve cellular lipids efficiently and are not able to keep proteins soluble enough under the conditions prevailing in IEF. To achieve both purposes, electrically neutral detergents are always added during sample preparation. Any type of detergent completely neutral over the whole pH range of isoelectric focusing and compatible with the chaotropes can be used. For practical reasons such as minimal interference with SDS-protein binding and minimal interference with protein assay methods, CHAPS is most often used [42] and is quite efficient in most cases. However, it is not the most efficient choice for difficult proteins (e.g. membrane proteins) and other detergents have been found to be much more efficient [43].

The second important point that has profound consequences is the fact that isoelectric focusing operates at very high field strengths. This is due to the fact that the charge of the proteins continuously decreases as they approach their pI. Thus, to make them travel the last millimeters close to the pI at a decent speed, very high electric fields must be used. Fields as high as 170V/cm are quite common in isoelectric focusing, compared to the 15V/cm commonly used in SDS electrophoresis.

This use of high field strengths imposes in turn to work at very low ionic strengths, because of Joule heating. This means in turn that it is very difficult to break ionic interactions during sample preparation for 2D gels. While protein-protein ionic interactions are usually conformation-dependent and therefore broken by protein denaturation (although there are unpleasant exceptions [44]), protein-nucleic acids interactions are much more difficult to break and give rise to insidious artifacts. One way to break them is to operate at high pH in the presence of a nucleic acid precipitant [45], and another way is to degrade the nucleic acids with trichloroacetic acid [46]. The latter paper, using mouse liver as the sample, shows that there is no universal way to prepare a perfect sample. For example, the use of trichloroacetic acid dramatically improves the extraction of basic, ribosomal proteins, but induces severe losses for several high molecular weight proteins.

Unfortunately, nucleic acids are not the sole type of interfering biological compounds. All polysaccharides with high charge density (ionic interactions), polyphenols able to crosslink proteins, or hydrophobic chemicals able to overcome the solubilizing power of detergents and chaotropes and to induce hydrophobic precipitation, just to quote a few, are interfering substances that must be eliminated. This process can be quite difficult for some biological samples that are quite rich in such substances, e.g. plant samples, for which dedicated protocols had to be devised [47, 48].

The last caveat in sample preparation concerns protein modification during the sample preparation. These can be chemical modifications, e.g. carbamylation induced by the presence of urea [49], and this is indeed more a problem during sample preparation and storage than during 2D electrophoresis itself [50]. However, artefactual modifications can also be brought by biological compounds, e.g. proteases remaining active during sample preparation. It is commonly assumed that the combination of chaotrope-induced protein denaturation and proteases inhibitors is sufficient to alleviate this problem, but depending on the sample this assumption can be completely wrong, in which case special denaturation processes must be used to destroy the proteases [51].

Thus, there are numerous protocols optimized for each type of sample, and it would be beyond the scope of this tutorial to give a comprehensive listing of sample preparation protocols. However, there are books dedicated to this topic [52, 53].

## 2.2. Isoelectric focusing

Compared to the delicate step represented by sample preparation, isoelectric focusing by itself is less problematic and generally much more straightforward, especially when immobilized pH gradients are used. There are however some points that deserve attention, i.e. sample application strategy, voltage profile, thiol ionization and isoelectric precipitation.

As isoelectric focusing is a steady-state technique, the sample can be theoretically applied at any point of the pH gradient without altering the final result. This is not true in practice, and the sample is generally applied at one of the extremities of the pH gradient or dispersed within the isoelectric focusing gel. In carrier ampholytes-based focusing, dispersion requires to include the sample in the gel during the polymerization of the gel itself [2, 3]. This is a rather cumbersome process, but it has been shown to increase the resolution of the resulting 2D maps by decreasing protein precipitation [54]. This process is much easier to achieve when immobilized pH gradients are used, as it is sufficient to include the sample in the strip rehydration solution [45]. While this technique is quite easy to use and allows to load important amounts of sample, it has been shown that it induces rather severe protein losses [55]. Furthermore, it fails completely when basic pH gradients are used, in which case application at the anodic side of the focusing gel remains mandatory (e.g. in [56]).

Voltage profile is a direct consequence of the high field strengths that must be used in isoelectric focusing. Quite often biological samples are not completely desalted, and the salts present can induce a strong Joule heating if a high field strength is applied at once. This is also true at a lesser extent for the ampholytes that are used as conductivity smootheners, which must reached their isoelectric point. For both salts and ampholytes, advantage is taken of their high mobility compared to proteins, and a moderate field strength is used for some time (typically 15V/cm for 3 hours) to allow them to reach their final position. Then the high field strength is applied. Special attention is brought to this point when samples containing a higher salt content than usual are present in a run, either alone or parallel to less salty samples. In both cases, it is advised to increase this low-field plateau phase, and never to use overall watt-control of the migration. If this is used, then the salt-rich samples will overheat compared to the low-salt samples, resulting in distorted migration.

Thiol ionization is a problem that occurs only at high pH. Cysteine thiols and tyrosine phenols ionize around pH 10. Due to their low abundance, and up to pH 8, their influence on the pI of the proteins is negligible. However, this is no longer true at pH 9 or higher. While phenol functions do not pose any problem, thiols do because of the thiol-disulfide exchange, as thiols ionize and disulfides do not. Thus keeping a constant redox status for cysteines is essential when isoelectric focusing at basic pH is performed. To keep a long story short, the best way to assure this and also to deal with cysteine protection is to convert the cysteines to disulfides [56, 57].

Isoelectric precipitation is the essential drawback of isoelectric focusing. It is really consubstantial to any separation where the proteins are separated at their pI, which means that it affects chromatofocusing as well as isoelectric focusing. This is linked to the fact that the isoelectric point is by definition a solubility minimum, especially at low ionic strength, as this is the exact point where no electrostatic repulsion will exist between protein molecules. By removing one of the most important molecular repulsive force, presence at the pI means extreme sensitivity to precipitation. This problem increases of course as the intrinsic protein solubility decreases, and this explains why 2D electrophoretic analysis of membrane proteins has proved so unsuccessful [58] and is very likely to remain so.

Finally, compared to the very early days where IEF was carried out only with carrier-ampholytes-driven pH gradients, it should be recalled that the now widespread use of immobilized pH gradients has made the whole 2D gel process much easier and much more performing. Besides the ease of use and end of deformation brought by the plastic-supported IPG strips, immobilized pH gradients have considerably improved reproducibility [26], but have also allowed pH gradient engineering, including non-linear pH gradients [59], basic gradients [60] or narrow pH gradients [61], which was very difficult to carry out with carrier ampholytes pH gradients.

## 2.3. Equilibration between dimension and SDS electrophoresis

The purpose of the inter-dimension equilibration process is to coat the proteins separated in the isoelectric focusing gel with SDS, so that they become mobile in the second dimension. This is always achieved by equilibration of the isoelectric focusing gel in a SDS-containing buffer [4], but it has been shown to be more delicate for immobilized pH gradient strips due to electroendosmosis problems [20, 21]. Moreover, the use of organic disulfides has further simplified the whole process [56].
However, it must be kept in mind that proteins, even if they are not truly precipitated, are likely to be somewhat insolubilized in the isoelectric focusing gel, especially when immobilized pH gradients are used, as suggested by results obtained on native proteins by zymograms [35]. Thus, it is wise to start the SDS electrophoresis at low voltage, just to give time to the SDS front to re-solubilize the proteins while sweeping across the immobilized pH gradient strip. Then SDS electrophoresis can be conducted the usual way.

## 2.4 Protein detection and image analysis

In the current proteomic landscape, 2D gel-based proteomics is the only setup in which there is an intermediate readout before the mass spectrometry stage, and this readout is precisely the on-gel detection of proteins. Thus, this step plays a crucial role, as i) only what is detected can be further analyzed and ii) quantitative variations observed at this stage are the basis to select the few spots of interest, in comparative studies, that will be the only ones processed for further analysis with mass spectrometry.

Consequently, there are enormous constraints and demands on this protein detection step, such as sensitivity, linearity and homogeneity of response, as well as compatibility with downstream processes such as mass spectrometry. It is therefore no surprise that this step is thoroughly described in the practical books devoted to 2D electrophoresis [32-34] and in recent reviews [37]. Here again, it is beyond the scope of this tutorial to go into the details of in-gel protein detection methods, but guidelines can be given. Current in-gel protein detection methods fall into three major categories: detection with organic dyes, silver staining and fluorescence.

Detection with organic dyes can be summarized in one single process, colloidal Coomassie Blue staining [62], which has really become a reference standard. Although the sensitivity is moderate, linearity and homogeneity are good and compatibility with mass spectrometry is excellent [63]. Conversely, silver staining is much more sensitive [64] but less linear and homogeneous, because of its delicate mechanism [65], and its compatibility with mass spectrometry is problematic [63]. As this has been shown to be a consequence of the presence of formaldehyde at the image development step [66], formaldehyde-free silver staining protocols have been recently proposed [67].

Finally, protein detection by fluorescence seems to be a happy compromise. It combines good sensitivity with excellent linearity, and compatibility with mass spectrometry is good [63]. In addition, several modes of detection can be used as reviewed in ( [68] ): environment-sensitive probes, non-covalent binding and covalent binding. The latter approach deserves some specific comments, as the use of chemically-related, reactive fluorescent probes differing mainly by their excitation and emission wavelengths allows to perform multiplexing of samples on 2D gels [69]. This multiplexing process solves in turn two difficult problems in the comparative analysis of gel images, namely the assignment of small positional differences and taking into account moderate quantitative changes [70-72]. Moreover, some variants of this process can be applied to very low amounts of samples [73].

However, it must be stressed that the image analysis process in 2D gel-based proteomics is usually not the simple comparison of two gels, but the multiple comparison of several gel images, since the very early days of the technique [13, 14]. In such a process, the real positional reproducibility of the gel is a crucial parameter. Although the immobilized pH gradients have dramatically increased this reproducibility [26], it has been recently demonstrated that reproducibility is always better when the gels are run in parallel [74], exactly as described more than 30 years ago [9, 10]. Last but certainly not least for this section, great care should be brought to experiment design and to statistical considerations to avoid falling into false positive issues when 2D gels are analyzed [75].

3. Current uses of 2D electrophoresis in proteomics.

Despite the now well known limitations of 2D gels, which have been outlined above for membrane proteins and will be dealt with in more detail in section 5, 2D gels are still widely used in proteomics, and this roots in several key features [76].

One of these features deals with the economy of proteomics. As shown on Figure 2, in 2D gel-based proteomics, the 2D gel part represents the essential workload of the whole process. It is at this step that the quantitative analysis is performed, and this quantitative analysis is usually used to perform spot selection. This has important consequences for the downstream mass spectrometry analysis. First of all, this means that only a very limited portion of the proteins present in the samples will need to be analyzed. This is especially true for a comparative study with replicates. If we imagine ten samples to be analyzed and compared, at 20 hours of mass spectrometry per sample, this represents in shotgun-type techniques 200 hours of MS. If the same analysis is carried out with 2D gels, at the end of the image analysis, maybe 20 different spots will be selected, and this represents at the very most 20 hours of mass spectrometry. This does not mean that 2D gel-based proteomics is more productive per se, it means that the burden put into the more expensive, MS part is reduced in this scheme. However, when analyzing the comparative productivity of 2D gel-based proteomics and shotgun-type proteomics, it appears that the compared productivity of 2D gels over shotgun improves when the size of the sample series increases, due to the highly parallel nature of 2D gel-based proteomics.

Second, and due to the high resolution of 2D gels, very simple and cheap MS process can be used to identify a protein from a 2D gel. For example, the old peptide mass fingerprinting method [77], which is fairly cheap, fast, and can be carried out on low-price TOF MS, works only with 2D gel-separated proteins, and will never work with any other technique of less resolving power. Furthermore, because of its low intrinsic consumption in mass spectrometer time, 2D gel-based proteomics can be carried out in a "hub and spokes" model, where several "peripheral" biology-oriented laboratories carry out all the 2D gel-based part of the proteomic analysis and come to the central mass spectrometry hub to carry out the last part. All in all this makes 2D gel-based proteomics quite efficient, economically speaking.

However, this is not the main reason of the popularity of 2D gel-based proteomics. The core reason resides into the reproducibility and robustness of the technique, as well as in the ease of the quantitative analysis. Indeed, quantitative analysis can be easily carried out on large series, i.e. tens of samples, well beyond the multiplexing capacities of mass spectrometry-based quantification. In fact, such a use of 2D gels to carry out multiple analyses on replicate samples is so common for this technique that the level of requirement for publication is much higher for gel-based proteomics than for shotgun-type proteomics [78, 79]. Consequently, 2D gel-based proteomics is widely used in areas where large series of samples are the norm, for example in toxicology (e.g. in [80] ).

At the other end of the spectrum, 2D gel-based proteomics is also widely used in bacterial proteomics, when the complexity of the sample is low enough to make the limits of gel-based proteomics less acute [81] .

There are also some niche applications where special traits of 2D gel-based proteomics are taken to profit. One example is micro enzymology, where 2D electrophoresis is used as an protein micropreparative tool [82]. Another example is immunoproteomics, where it is the immune response of patients that is probed at a proteomic level (e.g. in [83-85]). In this case, what is used is:

i) the ease of interface of 2D gels with antibody probing via the classical blotting process, and

ii) the fact that the resolution of 2D gels is such that in the identification stage, there is very little chance that the identified protein does not correspond to the immunodetected one, while the probability of comigration would be much higher in less resolutive systems.

Last but certainly not least, 2D gels are also very appropriate when post-translational modifications are studied. First, the same procedure of blotting can be used easily with antibodies directed against a modification, e.g. tyrosine nitration (e.g. [86]), citrullination [87] or hydroxynonenal adducts [88].

Second, many post translational modifications do alter the pI and/or the Mw of the proteins and thus induce position shifts in 2D gels. This is true for example for phosphorylation (e.g. as early as 1983 in [89]), glycosylation (e.g. in [90]), but also more delicate modifications such as glutathionylation [88], or more forgotten modifications such as protein cleavage (e.g. in [91]).
In addition, this feature of protein migration alteration upon modification can be used in a completely unsupervised scheme, where the modified protein is first detected as a spot with anomalous migration, and then the modification is identified and located by mass spectrometry techniques. Such as example is developed in the next section.

4. Worked example

The worked example is taken from a publication on oxidative stress response [92]. As oxidative stress is consubstantial to aerobic life, an induced oxidative stress will result into an exaggeration of the normal oxidative stress response, i.e. into quantitative variations. Moreover, direct protein modifications can be expected, either from oxidative modifications of proteins or from controlled post translational modifications (e.g. phosphorylation). Figure 3 shows the initial stage of the work, i.e. the pure 2D gel work. Quantitative changes of a few spots can be noted. Then mass spectrometry analyses showed that the changes concerned the same family of proteins, namely peroxiredoxins, and went by pairs (Figure 4), suggesting that the proteins were post-translationally modified upon oxidative stress. Then, detailed analysis carried out on the modified form of peroxiredoxin 2 demonstrated oxidation of the active site cysteine into cysteic acid (Figure 5) , thereby explaining the pI shift observed on 2D gels. However, it must be stressed that the identification of this modification required 50 pmoles of the modified protein, imposing to load 5 mg of total cell extract onto the 2D gels, and thus exploiting fully the micropreparative capacities of the technique.
Retrospective analysis explains why such an amount of protein was needed to discover the modification. As shown on Figure 6, trypsin digestion produces a heavy and hydrophobic peptide that contains the active site cysteine. Furthermore, this peptide gets an extra negative charge when the cysteine is oxidized. It is therefore no surprise that such a peptide has never been observed, and indeed it would be missed by any shotgun-type experiment. What was really observed was a peptide with two missed cleavages, generated because of the high amounts of protein digested. These two missed cleavages induce extra mass compared to the theoretical one, but also extra ionization sites, resulting in both improved water solubility and improved ionization.
This example shows how a simple spot wandering on a 2D gel can be drilled down to the site and nature of the unclassical post-translational modification conferring the altered migration. It also shows how the separative and preparative capacities of 2D gels can be used for such a purpose, thereby enabling the micropurification of the altered form. It also shows that sometime the modified form can be easily visualized at the protein level, although the modified peptide would be missed in any direct peptide-based proteomics scheme.

## 5. Current limitations and working limits

Because the technique is known for a very long time, the limits of 2D gels are also very well known. Beside the limit in the analysis of membrane proteins, which has been exposed in section 2 and is now very well documented [58], there is another very important limitation that is linked to the expression dynamics of proteins, especially in eukaryotic cells. A good illustrating example can be found with budding yeast. In *S. cerevisiae*, various studies have been made to determine at which level each gene is expressed, at a genome-wide scale. One set of data, obtained by GFP fusions for 4000 yeast genes [93], can be used as a starting point. When these data are compiled, the following figures are obtained: 130 gene products (i.e. 2% of the yeast genome) account for 50% of the protein mass. The top 10% most expressed genes (i.e. 400 genes in this experimental set) produce 75% of the protein mass. Conversely, the low expressed 2/3 of the genome (in this case 2/3 of 4000 tested genes i.e. 2500 genes) only produce 10% of the total protein mass. When these data are compared to the most advanced 2D gel-based yeast protein map [94], the latter shows identification for 485 gene products, i.e. the 10% most expressed genes only. When transposed to mammalian cells, where both the gene number and expression dynamics figures are even less favorable [95], 2D gels resolve a limited collection of highly abundant and soluble proteins, recently called "the déjà-vu in proteomics" [96], [97]. These proteins encompass a limited number of cell functions, mainly central metabolism, protein production (e.g. ribosomal proteins, some RNA biding proteins, translation factors), protein confomational control and degradation (chaperones, disulfide isomerases, proline isomerases, proteasome subunits), cytoskeleton at large (including some cytoskeleton modifying proteins), adaptor proteins (14-3-3, annexins) and oxidative stress response (catalase, superoxide dismutases, peroxiredoxins, glutathione transferases). All these proteins, with their major degradation products and post-translational variants, make the ca. 1000 protein spots that are usually seen on a 2D gel when a total cell extract of a mammalian cell is analyzed. This limited analysis scope of 2D gels explain why so many so-called "specific" markers of various cellular states, including disease state, all belong to the same classes [96], and correspond indeed to a core stress response module of the cells [97]. This does not mean that this core response is not important to the cells under the circumstances of interest, but it just means that the part of the response that is observed with gel-based proteomics is not specific in many cases. However, it is fair to say that these limitations, if documented recently, had been identified long before [98].

Thus, 2D gels must be seen as a protein screening process with high resolution, high reproducibility, quantitative, label-free intermediate read-out, but with limited analysis depth in terms of protein scope. Then, optimal use of 2D gels in proteomics is obtained when the complexity of the sample gets closer of the figures of merit of 2D gels in terms of proteins numbers and expression range. This occurs naturally when samples coming from less complex organisms are analyzed, such as bacterial samples [81, 95]. In these less complex organisms, transcriptional regulators are for example easily identified [99, 100] while they are completely out of reach in eukaryotic samples.

When dealing with eukaryotic samples, optimal use of 2D gels means to analyze not the complete cell, by far too complex, but only a fraction of it, which complexity will be lower and closer to the resolving power of 2D gels. Thus, the key word to get into less abundant cellular proteins is fractionation. Biochemical fractionation can be used [101], but an interesting trend consists in analyzing cell organelles, which complexity is much lower. As shown on the example of mitochondria [64], this gives access to low abundance, and sometimes poorly annotated proteins, while at the same time providing a reasonable evidence for their intracellular localization. This process is however not limited to intracellular proteins, as the analysis of secreted proteins has been shown to give access to low abundance proteins such as cytokines [102].

6. Future directions

With over 3 decades of research in the field, 2D electrophoresis is now a mature technique. It offers great flexibility and tunable resolution [103], and many tricks are available in the literature to tune and refine the resolution when needed (reviewed in [39]). There is therefore little gain to expect as to the pure performances of 2D gels by themselves.
When keeping to fully denaturing 2D electrophoresis, several systems purely based on zone electrophoresis have been developed, mainly to alleviate the solubility problem encountered with membrane proteins. While these systems did alleviate this specific problem [104], they never came to wide use because of their unsufficient resolution, thereby offering no real advantage over techniques based on 1D SDS PAGE [105].
Thus, progress in the optimal use of 2D gel-based proteomics will come from two sides. The upstream side will be to use more and more 2D gels on cell fractions, and decisive progress will come when more robust fractionation techniques than the current ones will become available. The downstream side, if we can call it this way, will be to use 2D gels where it brings most added value compared to other proteomics setups. One of these areas is clearly in the determination of post-translational modifications. Not only the list of modifications that can alter one protein, but also how these modifications are cooperative or mutually exclusive, and interplay to modulate the stability and/or function of cellular proteins. This knowledge begins to appear in the literature [106, 107], and 2D gels are a well-suited micropreparative tool for such studies [108, 109].
As a final remark, we would like to draw the attention of the reader on some figures related to the complexity of living organisms. C. elegans, a worm with around 1000 cells in total, has a genome that encodes 20,000 protein-coding genes producing ca. 25,000 different mRNAs. The human genome encodes 22,000 protein-coding genes, producing ca. 50,000 different mRNAs. Yet, a human body is made of 100,000 billion cells of 200 cell types, has a complex development and a highly complex central nervous system. This means that the difference in complexity between the worm and the human cannot be accounted for at the genome complexity level, and even not at the transcriptome complexity level, but only at the proteome level. And what is specific to the proteome level is on the one hand the amount of each protein, and on the other hand how protein functions of any type can be modulated by modifications. Phosphorylation is clearly not the only one, although it is the best-known one, and prenylation is just one other example of simple modification for which we know a role [110], not even to speak of the complexity of glycosylations and of their multiple roles.

Thus, the name of the game for proteomics in the future will be to determine quantities and qualities of proteins. As to determining protein quantities, there will be some competition coming from ribosome profiling [111], which will provide insights into protein synthesis rates. Of course these data are just indicative, but they can be reached rather easily with currently available technology. There will be however no match for proteomic techniques as far as post-translational modifications are concerned, and any tool that can be useful in this subfield of proteomics will prove useful for the future of proteomics. With its unique ability to separate at high resolution intact proteins with their associated collection of post-translational modifications [108, 109], it is quite clear that 2D gel-based proteomics will still be used in the 21st century proteomics.

References

* key major reviews
** key historical/breakthrough article
# reference book
##entry-level guides into the subject

proteins for internal sequence analysis after one- or two-dimensional gel electrophoresis. Anal Biochem. 1992;203:173-9.

dimensional gel electrophoresis system suitable for the separation of integral membrane proteins. Anal Biochem. 1996;240:126-33.

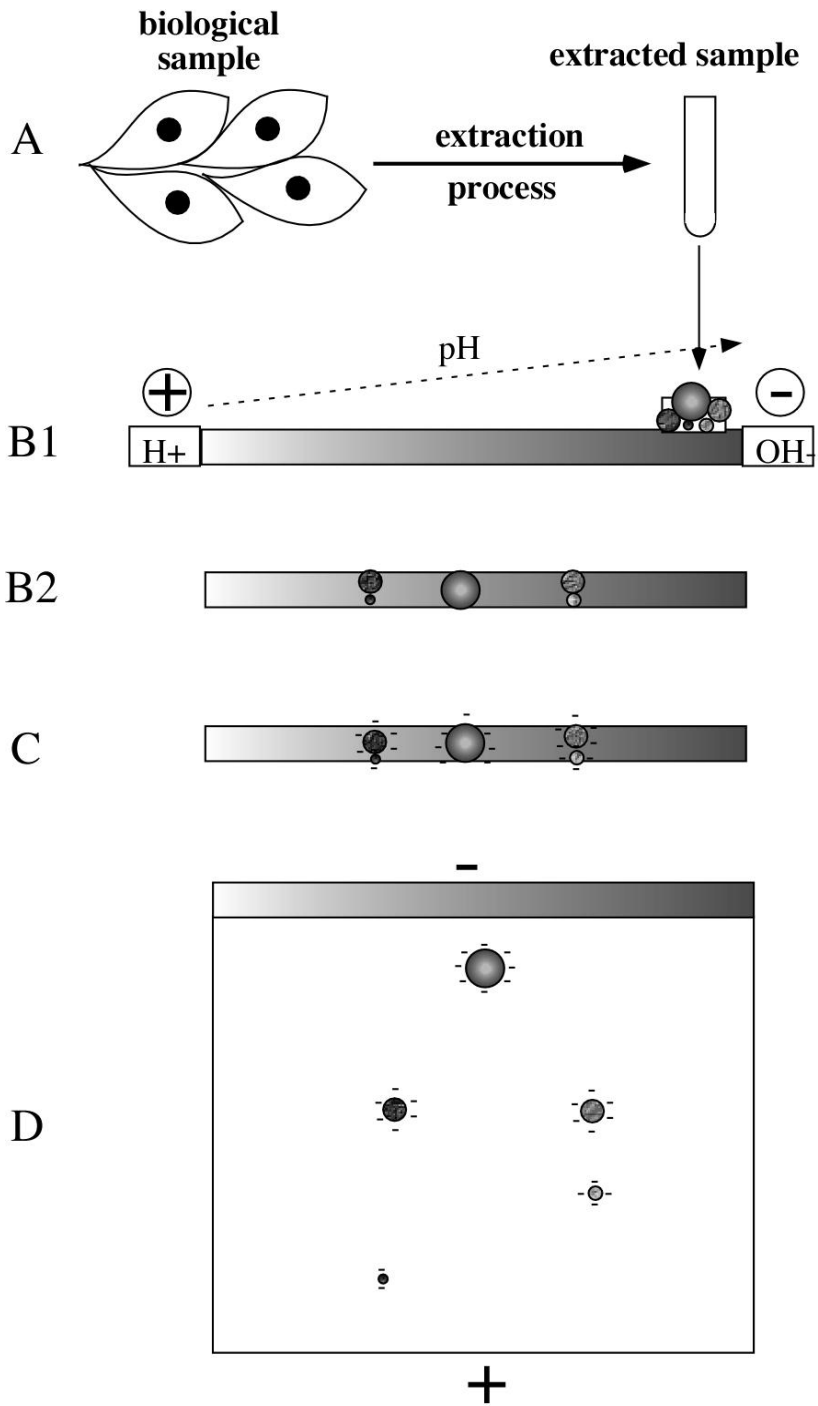

Figure 1: Scheme of principle of 2D gel electrophoresis
The total process start with the extraction of proteins from the biological sample to get an IEF-compatible sample (A). The sample is then loaded onto a pH gradient (B1) oriented with the acidic side at the anode and the basic side at the cathode. After the IEF step, the proteins have reached their pI and thus have no remaining electrical charge (B2). The strip is then equilibrated in a SDS-containing buffer, so that all proteins becomes strongly negatively charged (C). The IEF gel is then loaded on top of a SDS PAGE gel, and the proteins are separated according to their molecular masses (D). After this step, the proteins are detected directly  on the gel.

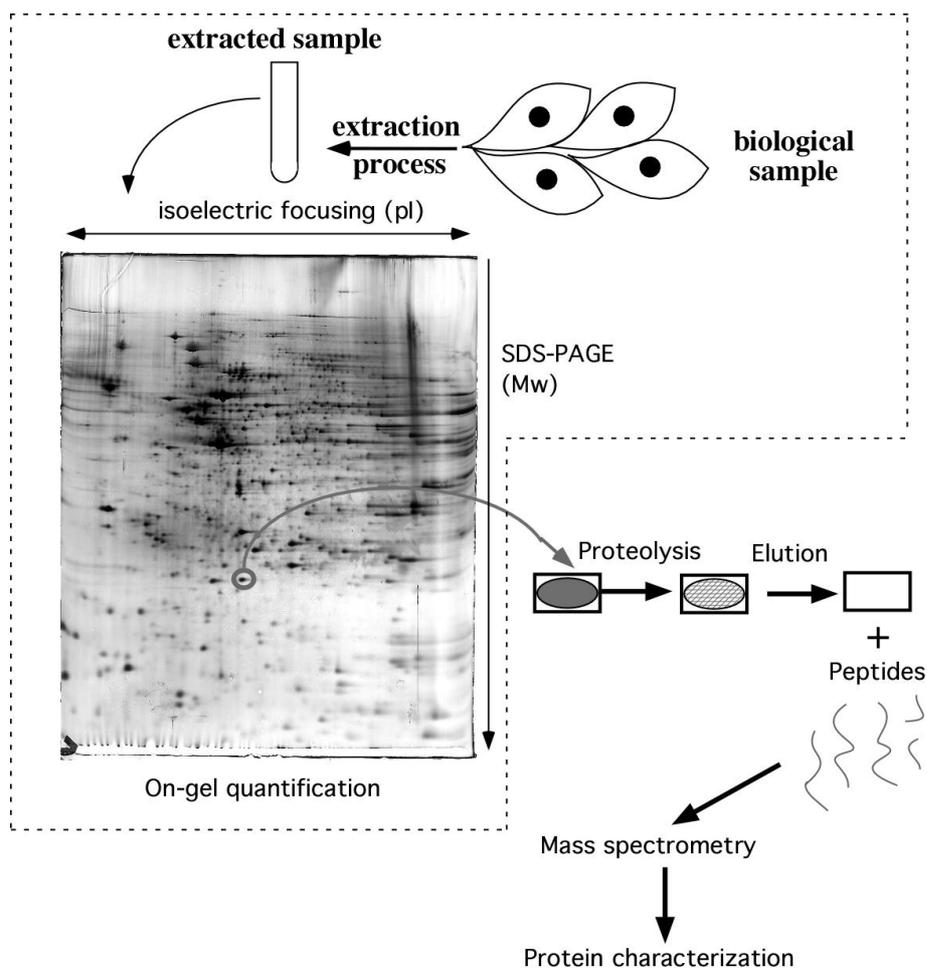

Figure 2: Scheme of principle of 2D gel-based proteomics
The first steps are exactly those described on figure 1. Then after detection of the proteins in the gels, the resulting images are quantitatively analyzed to determine the spots of interest. Those spots are then excised and submitted to in-gel digestion (generally with trypsin). The resulting peptides are then eluted and analyzed by mass spectrometry, leading to protein identification and characterization.
The dotted box shows the part of 2D electrophoresis in the whole process, and it can be easily seen that key steps, including sample preparation and quantitative analysis, take place during this procès

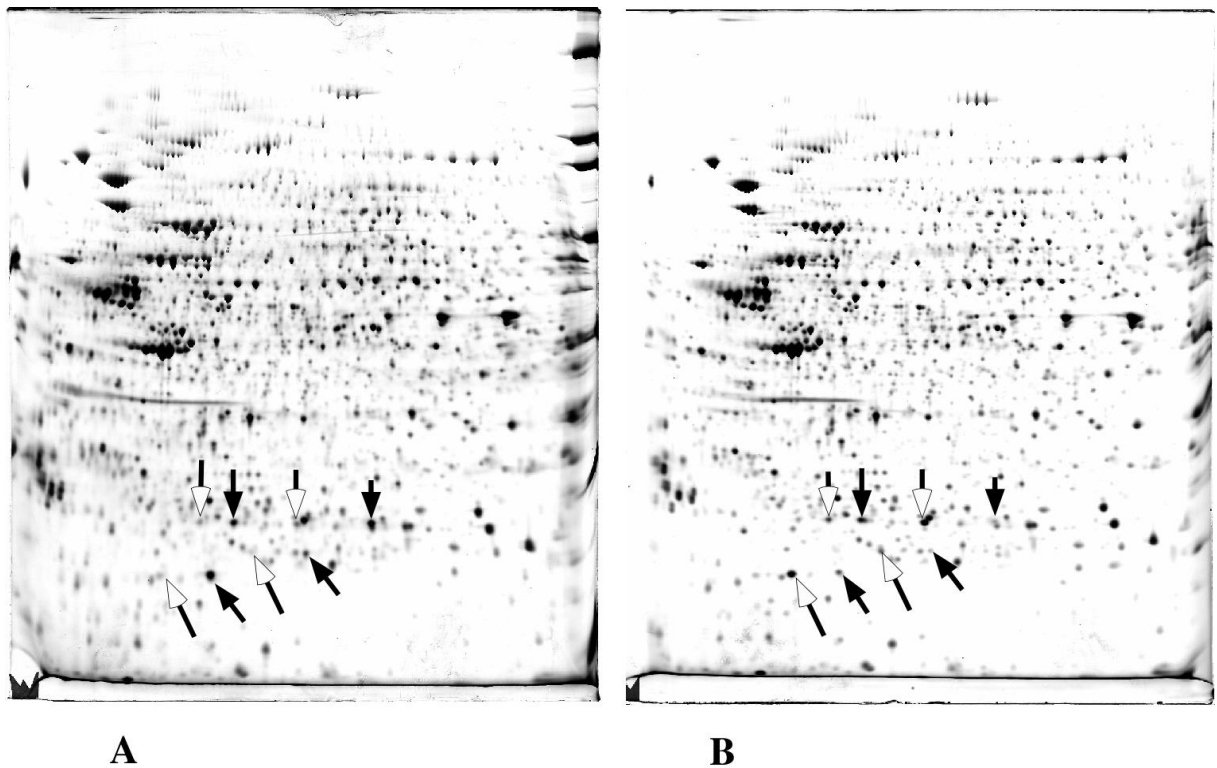

**A**                                                    **B**

Figure 3: First step of the analysis of cell response to oxidative stress.
Proteins extracted from control HeLa cells (A) and HeLa cells stressed with 0.15 mM
butylhydroperoxide for 2 hours (B) are separated by 2D gel electrophoresis. After
image analysis, only a few spots are found to change quantitatively between the two
situations. The spots more intense in the control cells are shown with a solid arrow,
and the spots more intense in the stressed cells are shown with a hollow arrow.
Figure adapted from the following original publication: Wagner et al. Biochemical
Journal (2002) 366: 777-785. © The Biochemical Society (with permission)

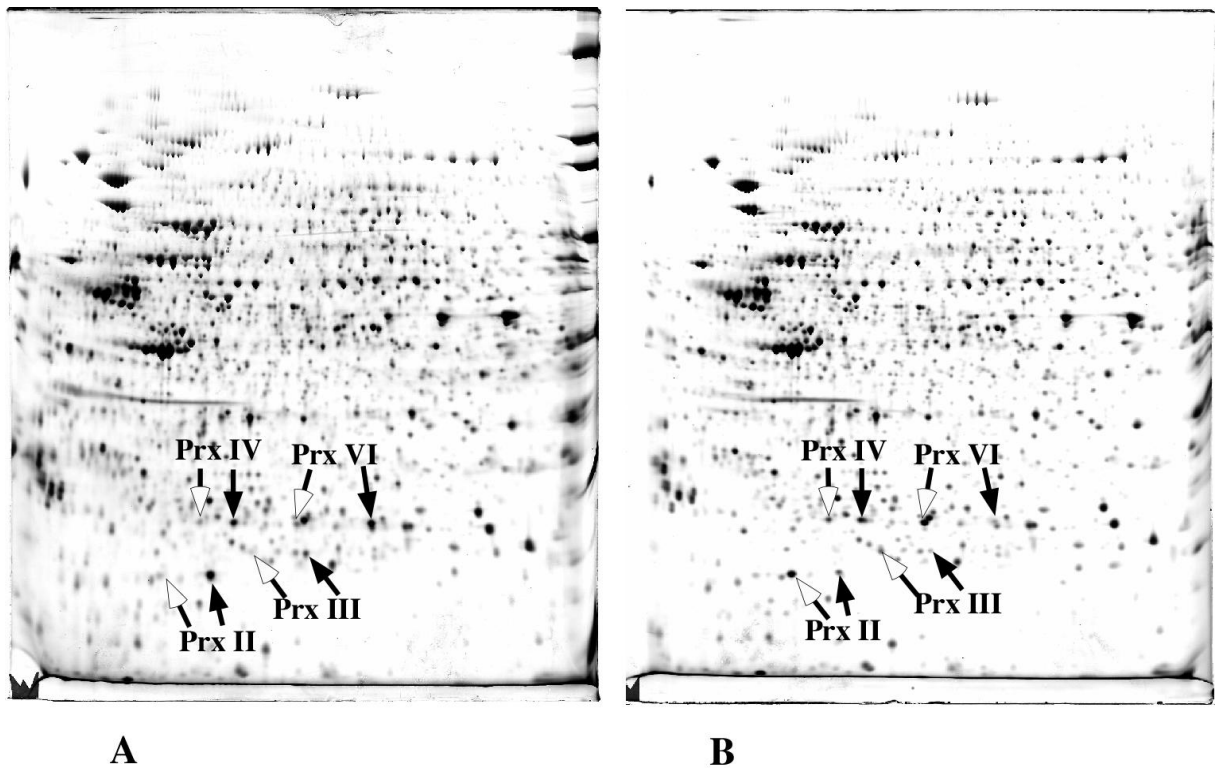

**A**           **B**

Figure 4: Second step of the analysis of cell response to oxidative stress
The spots of interest, determined by quantitative image analysis, have been submitted to the protein identification process with mass spectrometry, and the protein identifications are reported on the gels. It can be seen that spots go by pairs, with every time an important proportion of the protein relocalizing to a new position in the gel upon oxidative stress. This variation is typical of a massive post-translational modification of the proteins.
Figure adapted from the following original publication: Wagner et al. Biochemical Journal (2002) 366: 777-785. © The Biochemical Society (with permission)

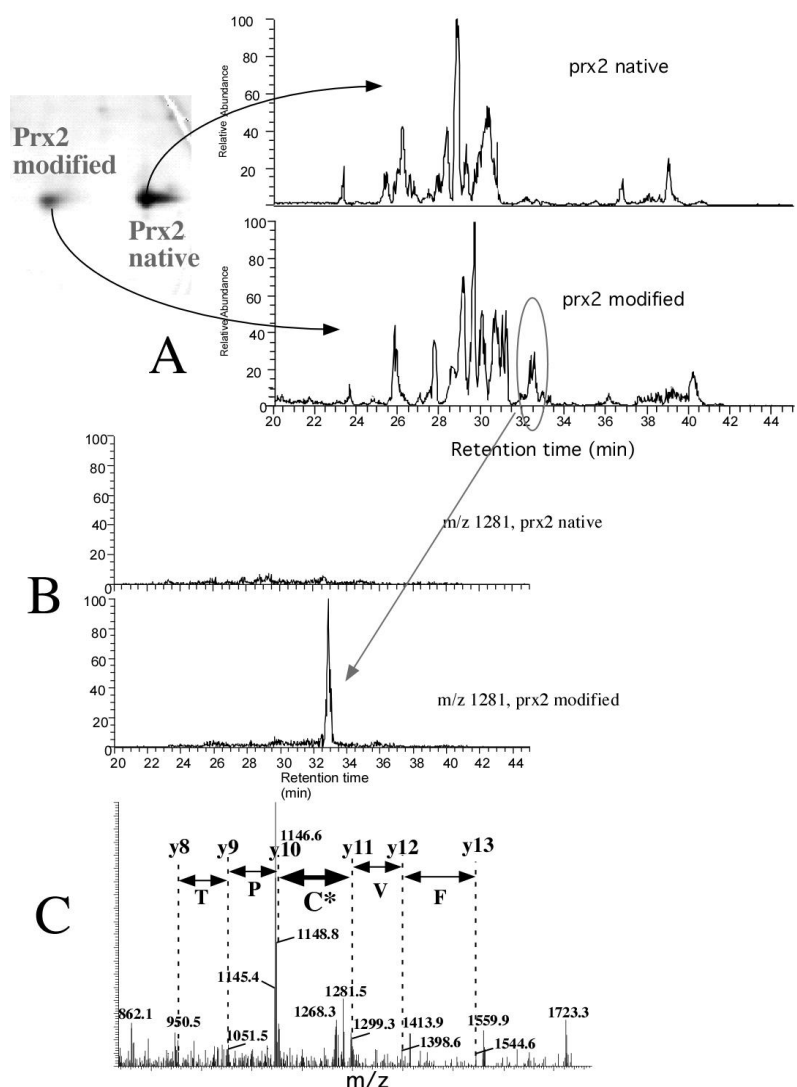

Figure 5: Determination of the post-translational modification on Prx2
Peroxiredoxin 2, one of the most abundant and easily-resolved peroxiredoxins, has been chosen to determine the post-translational modification taking place during oxidative stress. Both the "control" and "stress" form (ca. 50 pmoles each) are digested with trypsin, and analysed with LC/MS/MS (A). Comparative analysis of the two peptide profiles shows one peptide that is specific for the modified form (oval circled in the figure). This peptide is isolated on the fly (B) and fragmented by collision-activated dissociation. The fragment ions, shown in (C) allow to determine a partial amino acid sequence. Some of the fragments are separated by masses attributable to normal amino acids (shown is small font). A pair of fragments (thick double arrow) are separated by an abnormal mass, i.e. a mass that does not correspond to one of the 20 amino acids, but to a modified form of one of the 20 amino acids. The position of the fragments in the sequence allow to attribute the mass to a modified form of the active site cysteine (C* in large font on the figure), and the mass difference corresponds to the mass of cysteine + 48 Daltons, i.e. cysteic acid.
Figure adapted from the following original publication: Rabilloud et al. Journal of Biological Chemistry (2002) 277: 19396-19401. © The American Society for Biochemistry and Molecular Biology (with permission)

## PrxII (Swissprot P32119)

MASGNARIGK PAPDFKATAV VDGAFKEVKL SDYKGKYVVL FFY

PLDFTFV **C**PTEIIAFSN RAEDFRKLGC EVLGVSVDSQ FTHLAWINTP

RKEGGLGPLN IPLLADVTRR LSEDYGVLKT DEGIAYRGLF IIDGKGV

LRQ ITVNDLPVGR SVDEALRLVQ AFQYTDEHGE v**C**PAGWKPGS DTI

KPNV̶D̶D̶S̶ ̶K̶E̶Y̶F̶S̶K̶H̶N̶ - - - - - - - - - - - - - - - - - - - - - - - - - - - - -

## PrxII tryptic peptides

| | | |
|---|---|---|
| 3731.765 | 158-191 | LVQAFQYTDEHGEV**C**PAGWK PGSDTIKPNVDDSK |
| ***3001.52*** | *37-61* | **YVVLFFYPL**DFTFV**C**PTEII AFSNR |
| 2642.339 | 68-91 | LGCEVLGVSVDSQFTHLAWI NTPR |
| 1734.975 | 93-109 | EGGLGPLNIPLLADVTR |
| 1211.674 | 140-150 | QITVNDLPVGR |
| 1023.536 | 111-119 | LSEDYGVLK |
| 978.525 | 17-26 | ATAVVDGAFK |
| 972.551 | 8-16 | IGKPAPDFK |
| 924.442 | 120-127 | TDEGIAYR |
| 862.503 | 128-135 | GLFIIDGK |
| 789.410 | 151-157 | SVDEALR |
| 706.330 | 1-7 | MASGNAR |
| 673.319 | 192-196 | EYFSK |
| 637.294 | 62-66 | AEDFR |
| 625.319 | 30-34 | LSDYK |

- - - - - - - - - - - - - - - - - - - - - - - - - - - - -

## Observed modified peptide

L_SDY_K_G_K_YVVLFFYPLDFTF_ v*C*\*_PTEIIAFSN_R

Figure 6: Peptide maps of peroxiredoxin 2
On top of the figure, the sequence of the protein is shown. In the central part, the theoretical tryptic peptide map is shown, and the peptide containing the active site cysteine is boxed. In bold in the sequence, a tract of hydrophobic amino acids, a known factor for poor peptide extraction. Also note the large mass of the peptide (in bold italic) another factor for poor extraction and fragmentation. Because of these two poor prognostic factors, this peptide was never observed in our experiments. Even worse, cysteine oxidation introduces a negative charge, thereby further impairing peptide ionization in the positive mode.
In the bottom part of the figure, the sequence of the peptide actually observed on Figure 5 is shown. Because of the high protein amounts used, there was two cleavage sites (the two lysines indicated in large font) that were missed during trypsin digestion. These two sites conferred extra hydrophilicity close to the N-terminus, resulting in better peptide extraction. The residues contributing to the peptide charge are indicated in large font. The oxidized cysteine (in italics) accounts for one negative charge. The N-terminal leucine accounts for one positive charge, and the three basic amino acids (the two internal lysines and the C-terminal arginine) account for three positive charges. This makes a 3+ charged peptide of 3843 Daltons, corresponding to the 1281 m/z observed in Figure 5.